\documentstyle[12pt]{article}
\newlength{\abstractwidth}
\renewcommand{\thefootnote}{\fnsymbol{footnote}}
\renewcommand{\thanks}[1]{\footnote{#1}} 
\newcommand{\starttext}{
\setcounter{footnote}{0}
\renewcommand{\thefootnote}{\arabic{footnote}}}
\newcommand{\be}{\begin{equation}}
\newcommand{\bea}{\begin{eqnarray}}
\newcommand{\eea}{\end{eqnarray}}
\newcommand{\beq}{\begin{equation}}
\newcommand{\ee}{\end{equation}}
\newcommand{\eeq}{\end{equation}}

\newcommand{\<}{\langle}

\newcommand{\PSbox}[3]{\mbox{\rule{0in}{#3}\includegraphics{#1}\hspace{#2}}}
\renewcommand{\>}{\rangle}
\def\ba{\begin{eqnarray}}
\def\ea{\end{eqnarray}}

\begin{document}

\begin{titlepage}
\bigskip
\hskip 3.7in\vbox{\baselineskip12pt
\hbox{MIT-CTP-2770}\hbox{hep-th/9808006}}
\bigskip\bigskip\bigskip\bigskip

\centerline{\large \bf
Comments on 4--point functions in the CFT/$AdS$ correspondence}

\bigskip\bigskip
\bigskip\bigskip

\centerline{ Daniel Z. Freedman$^{a,b}$, 
Samir D. Mathur$^{a}$, }
\vskip .1in
\centerline{Alec Matusis$^{a}$ 
and Leonardo Rastelli$^a$}
\bigskip\bigskip\bigskip
\centerline{$^a$ \it Center for Theoretical Physics}
\centerline{ \it Massachusetts Institute of Technology}
\centerline{ \it Cambridge, {\rm MA}  02139}
\medskip
\centerline{$^b$ \it Department of Mathematics}
\centerline{ \it Massachusetts Institute of Technology}
\centerline{\it Cambridge, {\rm MA} 02139}
\bigskip\bigskip\bigskip\bigskip

\begin{abstract}
We study the four--point function  of chiral primaries corresponding to the
dilaton--axion sector in supergravity in the $AdS_5$/CFT$_4$ correspondence.
 We find relations between some of the supergravity graphs and compute
 their leading singularities. We discuss the issue of logarithmic singularities
and their significance for the OPE structure of the CFT.

\baselineskip=16pt

\end{abstract}
\end{titlepage}
\starttext
\baselineskip=18pt
\setcounter{footnote}{0}


\section{Introduction}

Recently it has been proposed that IIB string theory on $AdS_5\times S^5$ is
 dual to a CFT: ${\cal N}=4$ supersymmetric $SU(N)$ Yang-Mills on the boundary of $AdS_5$~\cite{maldacena,polyakov,witten}. Using this correspondence the 2-point functions~\cite{polyakov}--\cite{us} and 3-point 
functions~\cite{us}--\cite{dhoker} of  primary operators in the CFT 
were computed 
in the limit $N\rightarrow\infty$, $g_{YM}\rightarrow 0$, 
$g_{YM}^2N\rightarrow\infty$, where the string theory computations reduce to tree
 level supergravity calculations.

Several interesting physical issues  arise when we move to the study of 4-point functions.
 We will focus on the limit $N\rightarrow\infty$, $g_{YM}\rightarrow 0$,
 $g_{YM}^2N\rightarrow\infty$ mentioned above. In the CFT the scaling
 dimensions of the chiral primary operators (and their superconformal
 descendents) are protected, while the dimensions of fields corresponding
 to massive string states are infinite in this limit. Does there exist a `complete'
 set of fields and an operator product expansion (OPE) structure that allows
 us to obtain 4-point functions much the same as in the case of 2-D CFT? If
 so, do the chiral primaries and their descendents form the complete set, or do
 we need other fields in the CFT? Is there a connection between supergravity
 fields propagating in the internal leg of a supergravity graph, and the contribution
 of a specific chiral primary (plus descendents) in the OPE expansion of the
 corresponding CFT correlator? Preliminary
 results on these questions were 
presented in \cite{talk} and \cite{tseytlin}.

To address such issues we study in this letter some simple supergravity graphs
 corresponding to 4-point functions in the CFT. We consider the dilaton ($\phi$)
 and axion ($C$) sector. (This sector has also been 
 studied in \cite{tseytlin}, and, while we use similar methods, we arrive at somewhat different
 conclusions).

\section{4-point functions in the dilaton-axion sector}

The relevant part of the $AdS_5\times S_5$ supergravity action is
\bea
S&=&{1\over 2\kappa^2}\int_{AdS_5} d^5x \sqrt{g}[-R+{1\over 2}(\partial\phi)^2+{1\over 2}
e^{2\phi}(\partial C)^2]\nonumber\\
&=&{1\over 2\kappa^2}\int_{AdS_5} d^5x \sqrt{g}[-R+{1\over 2}(\partial\phi)^2+{1\over 2}
(\partial C)^2+a\phi(\partial C)^2+b\phi^2 (\partial C)^2+\dots]
\label{one}
\eea
where $a=1, b=1$.
We use coordinates where the (Euclidean) $AdS$ space appears as the upper half space
($z_0>0$) with metric:
\be
ds^2={1\over z_0^2}[dz_0^2+\sum_{i=1}^d dx_i dx_i]
\label{two}
\ee
The $AdS$ space has dimension $d+1$; thus in our present case $d=4$.

First consider the CFT correlator $\<O_\phi(x_1)O_C(x_2)O_\phi(x_3)O_C(x_4)\>$. In
 the AdS calculation we encounter the supergravity graphs shown in Figure 1. 
\begin{figure} 
\begin{center} \PSbox{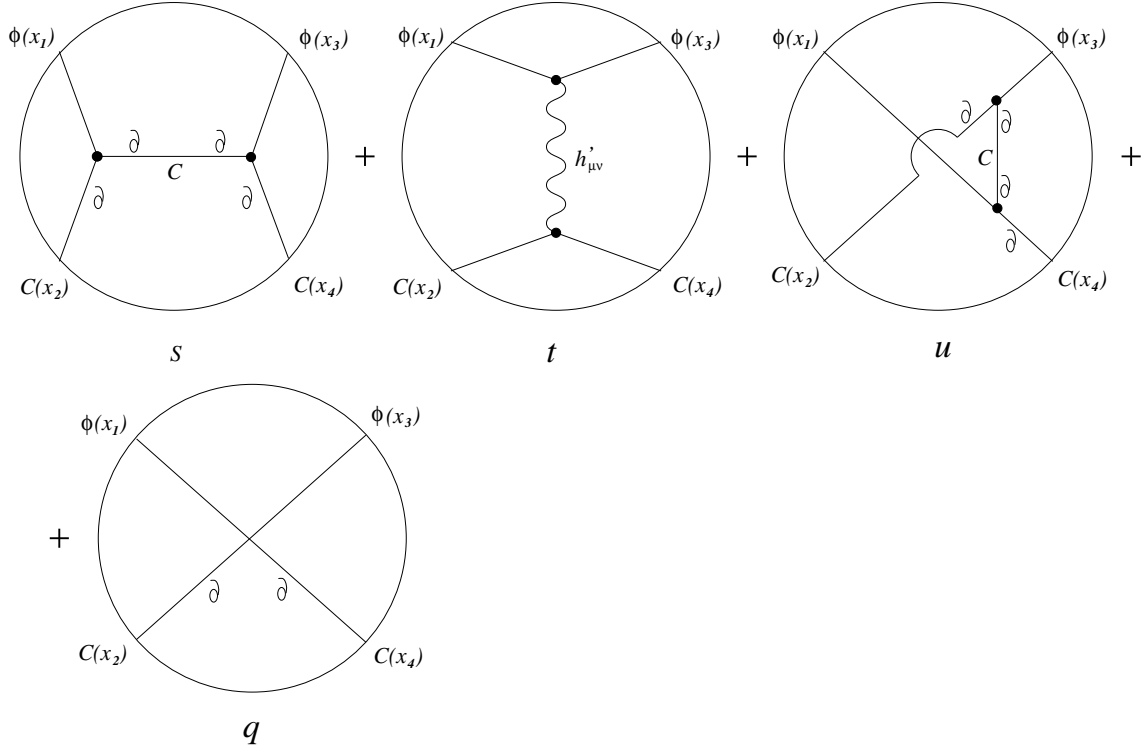 hscale=80 vscale=80}{5.8in}{3.5in} \end{center}
\caption{Supergravity graphs contributing to $\<O_\phi(x_1)O_C(x_2)O_\phi(x_3)O_C(x_4)\>.$}
\end{figure}
The s-channel amplitude is
\be
s=-(4a^2)I^s_{\phi C\phi C}(x_1,x_2,x_3,x_4)
\label{three}
\ee
\bea
&&I^s_{\phi C\phi C}(x_1,x_2,x_3,x_4)\equiv\nonumber\\
&&~~~~~\int {d^5z\over z_0^5}{d^5w\over w_0^5}z_0^2w_0^2K(z,x_1)\partial_{z_\mu}K(z, x_2)\partial_{z_\mu}\partial_{w_\nu}G(z,w)K(w,x_3)\partial_{w_\nu}K(w,x_4)
\label{four}
\ea
where\footnote{We assume $\Delta>d/2$. The case $\Delta=d/2$ saturates the unitarity
 bound and requires a special normalisation\cite{us}.}
\be
K_\Delta(z,x)={\Gamma(\Delta)\over \pi^{d/2}\Gamma[\Delta-d/2]}\left({z_0\over z_0^2+
(\vec z-\vec x)^2}\right)^\Delta
\label{five}
\ee
is the normalized boundary to bulk propagator for scalar fields in  supergravity
 corresponding to  primary operators in the CFT of scaling dimension 
$\Delta$~\cite{witten,us}. We have $d=4$ and note that for both $\phi$ and $C$ we have
 $\Delta=4$. For this case we will simply write $K$ without subscript.  $G(z,w)$ is
 the bulk to bulk propagator in the $AdS_5$ space for massless scalar fields, satisfying
 \footnote{In \cite{tseytlin} 
the notation is instead $\triangle_zG(z,w)=-\delta(z,w)$.}
\be
\triangle_z G(z,w)=\delta(z,w)
\label{six}
\ee
We will not need the explicit form of $G(z,w)$.

The quartic graph is
\be
q=-(4b)I^q_{\phi C\phi C}(x_1,x_2,x_3,x_4)
\label{seven}
\ee
\ba
&&I^q_{\phi C\phi C}(x_1,x_2,x_3,x_4) \equiv \nonumber \\
&&~~~~\int {d^5z\over z_0^5}z_0^2K(z,x_1)\partial_{z_\mu}K(z, x_2)K(z,x_3)
\partial_{z_\mu}K(z,x_4)
\label{eight}
\ea

The combinatoric factors in (\ref{three}), (\ref{seven}) can be obtained either from
 Feynman perturbation theory of supergravity or directly from  the fourth 
variation of the supergravity action (\ref{one}) with respect to boundary values of the fields.

In \cite{tseytlin} a nice manipulation was given which relates $I^s$ to a 4-point contact graph:
\ba
&&\int {d^5z\over z_0^5}{d^5w\over w_0^5}z_0^2w_0^2K(z,x_1)\partial_{z_\mu}K(z, x_2)\partial_{z_\mu}\partial_{w_\nu}G(z,w)K(w,x_3)\partial_{w_\nu}K(w,x_4)\nonumber \\
&&=\int {d^5z\over z_0^5}{d^5w\over w_0^5}z_0^2w_0^2\partial_{z_\mu}K(z,x_1)K(z, x_2)\partial_{z_\mu}\partial_{w_\nu}G(z,w)K(w,x_3)\partial_{w_\nu}K(w,x_4)\nonumber \\
&&={1\over 2}\int {d^5z\over z_0^5}{d^5w\over w_0^5}z_0^2w_0^2
\partial_{z_\mu}[K(z,x_1)K(z, x_2)]\partial_{z_\mu}\partial_{w_\nu}G(z,w)K(w,x_3)\partial_{w_\nu}K(w,x_4)\nonumber \\
&&={1\over 2}\int {d^5z\over z_0^5}{d^5w\over w_0^5}w_0^2K(z,x_1)K(z, x_2)\delta(z,w)\partial_{w_\nu}K(w,x_3)\partial_{w_\nu}K(w,x_4)\nonumber \\
&&={1\over 2}\int {d^5z\over z_0^5}z_0^2K(z,x_1)K(z, x_2)\partial_{z_\nu}K(z,x_3)
\partial_{z_\nu}K(z,x_4)
\ea
where we have integrated by parts (noting that surface terms vanish), used the
 fact that $\triangle_zK(z,x)=0$,  and used (\ref{six}). Thus we see that
\be
I^s_{\phi C\phi C}(x_1,x_2,x_3,x_4)={1\over 2}I^q_{\phi \phi CC}(x_1,x_2,x_3,x_4)
\label{el}
\ee
\be
I^u_{\phi C\phi C}(x_1,x_2,x_3,x_4)={1\over 2}I^q_{C\phi \phi C}(x_1,x_2,x_3,x_4)
\label{tw}
\ee

Note that the RHS of (\ref{el}) or (\ref{tw}) is not the same as the quartic graph in Figure
 1(q) since the derivatives act on different variables.

It is easy to see by using integration by parts that
\be
I^q_{\phi \phi CC}(x_1,x_2,x_3,x_4)=I^q_{CC\phi \phi }(x_1,x_2,x_3,x_4)
\label{nine}
\ee
\be
I^q_{\phi \phi CC}(x_1,x_2,x_3,x_4)+I^q_{\phi C\phi C}(x_1,x_2,x_3,x_4)+
I^q_{C\phi \phi C}(x_1,x_2,x_3,x_4)=0
\label{ten}
\ee
Thus we find that the contributions to 
$\<O_\phi(x_1)O_C(x_2)O_\phi(x_3)O_C(x_4)\>$ from the s, u and quartic graphs add up to
\ba
&&-4a^2{1\over 2}I^q_{\phi \phi CC}(x_1,x_2,x_3,x_4)-4a^2{1\over 2}I^q_{C\phi \phi C}(x_1,x_2,x_3,x_4)-4bI^q_{\phi C \phi C}(x_1,x_2,x_3,x_4)\nonumber \\
&&~~~~~=(-4b+2a^2)I^q_{\phi C\phi C}(x_1,x_2,x_3,x_4)
\ea
Putting $a=1, b=1$ we see that the coefficient on the RHS is not zero. 
 In the next section we show that  the function $I^q_{\phi C \phi C}(x_1,x_2,x_3,x_4)$
 is nonzero
 by computing its leading singularities.

The 4-point function of  the primary operator corresponding to the axion field \\$\<O_C(x_1)O_C(x_2)O_C(x_3)O_C(x_4)\>$ is given by the $AdS$ graphs in Figure 2.
\begin{figure} 
\begin{center} \PSbox{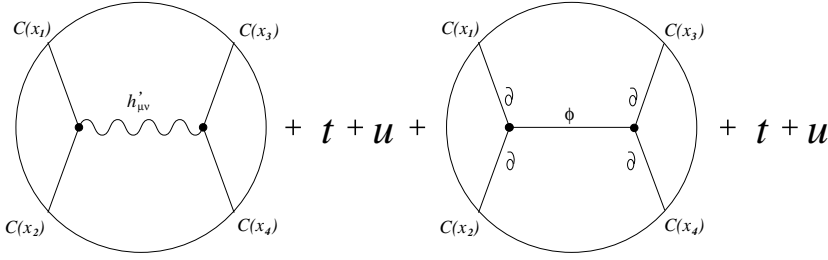 hscale=65 vscale=65}{5.6in}{1.5in} \end{center}
\caption{Supergravity graphs contributing to $\<O_C(x_1)O_C(x_2)O_C(x_3)O_C(x_4)\>.$}
\end{figure}
 Using (\ref{ten}) we see that the sum of the three dilaton exchange graphs sums to zero,
 though each of these graphs will not separately vanish.

\section{Singularities in 4-point graphs}

We have seen that the s and u graphs of Figure 1 reduce to the form of an $I^q$ integral.
 In the function $I^q_{\phi \phi CC}(x_1,x_2,x_3,x_4)$ there are two independent short
 distance limits to be considered:

(a) $x_{12}\equiv |x_1-x_2|\rightarrow 0$.

(b) $x_{13}\equiv |x_1-x_3|\rightarrow 0$.

(From (\ref{nine}) we see that $x_{34}\rightarrow 0$ is similar to $x_{12}\rightarrow 0$ etc.).

We first observe the identity
\ba
&&\int {d^{d+1}z\over z_0^{d+1}}z_0^2K_{\Delta_1}(z,x_1)K(z, x_2)_{\Delta_2}\partial_{z_\mu}K(z,x_3)_{\Delta_3}\partial_{z_\mu}K(z,x_4)_{\Delta_4} \\
&&~~~~=\Delta_3\Delta_4J_{\Delta_1,\Delta_2,\Delta_3,\Delta_4}(x_1,x_2,x_3,x_4)\nonumber \\
&&~~~~~~~~-2(\Delta_3-{d\over 2})(\Delta_4-{d\over 2})x_{34}^2J_{\Delta_1,\Delta_2,\Delta_3+1, \Delta_4+1}(x_1,x_2,x_3,x_4) \nonumber
\label{thir}
\ea
where
\be
J_{\Delta_1,\Delta_2,\Delta_3,\Delta_4}(x_1,x_2,x_3,x_4)\equiv
\int {d^{d+1}z\over z_0^{d+1}}K_{\Delta_1}(z,x_1)K(z, x_2)_{\Delta_2}K(z,x_3)_{\Delta_3}K(z,x_4)_{\Delta_4}
\label{fourt}
\ee
This identity can be derived by methods similar to those in \cite{us} (translating $x_3$
 to the origin, performing an inversion $z_\mu= \frac{z'_\mu}{(z')^2}, x_i =\frac{x'_i}{(x')^2}$,
 evaluating the derivatives and inverting back).

This manipulation reduces the calculation of an integral of the type $I^q$ to computing the
 quartic graph with no derivatives on any of the legs. A special case of this latter calculation
 (with all $\Delta_i=\Delta$) was given in 
\cite{canadians1}; we make a straightforward extension of their calculation to the case
 with arbitrary $\Delta_i$:
\ba
&&J_{\Delta_1,\Delta_2,\Delta_3,\Delta_4}(x_1,x_2,x_3,x_4)=\nonumber \\
&&{1\over 2\pi^{3d/2}}
{\Gamma[-{d\over 2}+{\sum_i\Delta_i\over 2}]\Gamma[-\Delta_4+{\sum_i\Delta_i\over 2}]\Gamma[-\Delta_3+{\sum_i\Delta_i\over 2}]\Gamma[\Delta_3]\Gamma[\Delta_4]\over \Gamma[{\sum_i\Delta_i\over 2}]\prod_i\Gamma[\Delta_i-{d\over 2}]}\nonumber \\
&&\int_0^\infty {d\beta_2\over \beta_2}(\beta_2 x_{24}^2+x_{14}^2)^{-\Delta_4}(\beta_2x_{12}^2)^{\Delta_4-\sum_i{\Delta_i\over 2}}
\left({x_{34}^2\over (\beta_2x_{24}^2+x_{14}^2)}\right)^{-\Delta_3}\beta_2^{\Delta_2}
\nonumber \\
&&{}_2F_1[-\Delta_4+\sum_i{\Delta_i\over 2}, \Delta_3, \sum_i{\Delta_i\over 2}, 1-\alpha]
\label{sixt}
\ea
where
\be
\alpha={(\beta_2x_{23}^2+x_{13}^2)(\beta_2x_{24}^2+x_{14}^2)\over \beta_2 x_{12}^2x_{34}^2}
\ee
and ${}_2F_1$ is the hypergeometric function.
For the estimates below it is helpful to use the integral representation:
\be
{}_2F_1[\alpha,\beta;\gamma, z]={1\over B[\beta,\gamma-\beta]}\int_0^1 t^{\beta-1}(1-t)^{\gamma-\beta-1}(1-tz)^{-\alpha}dt
\label{fift}
\ee
where $B[\alpha,\beta]$  is the Beta function.

${}$From (\ref{sixt}) and (\ref{fift}) we find that as $x_{12}\rightarrow 0$:
\be
I^q_{\phi \phi CC}(x_1,x_2,x_3,x_4)\rightarrow
{6^4\over \pi^6}{4\over 21}{1\over x_{13}^8x_{14}^8}\ln{x_{13}x_{14}\over x_{12}^2}
\label{eit}
\ee
As $x_{13}\rightarrow 0$:
\be
I^q_{\phi \phi CC}(x_1,x_2,x_3,x_4)\rightarrow
-{6^4\over \pi^6}{2\over 21}{1\over x_{12}^8x_{14}^8}\ln{x_{12}x_{14}\over x_{13}^2}
\label{sevt}
\ee

Note that the strengths of the singularities in (\ref{eit}) and (\ref{sevt}) 
are such that they
 respect the identity (\ref{ten}).

In \cite{tseytlin} it was argued that each of the s,u and quartic graphs 
given in Figure 1
 vanishes separately, while we have reached a somewhat different
 conclusion.\footnote{The resubmitted version (v4) of \cite{tseytlin} 
appears to agree with our
   conclusions.
} 
 We have not evaluated the graviton exchange graph, which was speculated 
to vanish in
 \cite{tseytlin}, but we discuss in the next section  our expectations
 for its contribution.

\section{Discussion}
We know that the ${\cal N}=4$ SYM theory is exactly conformal. Consider
a 4-point function $\< O_1(x_1)O_2(x_2)O_3(x_3)O_4(x_4) \>$
in the limit $x_1 \rightarrow x_2$, $x_3 \rightarrow x_4$. We might 
try to expand\footnote{See also \cite{ferrara} for 
 discussions of 
conformal OPEs and the  the contribution
of a given primary operator and its descendents to the CFT 4-point function.}
\beq
\label{ope}
O_1(x_1)O_2(x_2) = \sum_{n} \frac{\alpha_n O_n(x_1)}{(x_1 - x_2)^{\Delta_1 +
\Delta_2 - \Delta_n}}
\ , \ 
O_3(x_3)O_4(x_4) = \sum_{m} \frac{\beta_m O_m(x_3)}{(x_3 - x_4)^{\Delta_3 +
\Delta_4 - \Delta_m}}
\label{twenty}
\eeq
and get
\beq
\label{doubleope}
\< O_1(x_1)O_2(x_2)O_3(x_3)O_4(x_4) \> =
\sum_{n,m} \frac{\alpha_n \beta_m \, \< O_n(x_1) O_m(x_3) \>}{
(x_1 - x_2)^{\Delta_1+\Delta_2 - \Delta_m}(x_3 - x_4)^{\Delta_3+\Delta_4 - \Delta_n}}
\eeq
In a non-conformal theory, where a mass scale $m$ would be available,
we could also have, for instance, $O_{\Delta_1}(x_1) O_{\Delta_2}(x_2) 
\sim {\rm log}(m|x_1 - x_2|) O_{\Delta_1 + \Delta_2}(x_1)$, but in a conformal theory
such a term should not arise. Thus if the sums in (\ref{doubleope})
are to converge, we expect that the limit $x_{12} \rightarrow 0$ in
the correlator would have no term in ${\rm log}(x_{12})$. Individual graphs
from supergravity, however, are generically  expected to have such 
logarithmic singularities
and (\ref{eit}),(\ref{sevt}) are examples of this fact. Thus either the logs
 all cancel when the 
supergravity graphs are summed, or a naive OPE summation of the form (\ref{doubleope})
is invalid. 

We now proceed to discuss our results for 4-point functions in the 
dilaton-axion sector
in the light of the questions of cancellation of logs and expectations
 for power singularities.
For the correlator $\< O_{\phi} O_CO_{\phi}O_C\>$ we found in (14) that
 the sum of s,u and
 quartic graphs
is proportional to the contact amplitude and contains logarithmic 
singularities. We have not
 evaluated
the t-channel graviton exchange graph, which is quite difficult, but
 which could contain 
logarithms that cancel those in the sum s+u+quartic. Note that
if such a cancellation occurs for the $AdS_5\times S_5$ supergravity 
theory then it would
 certainly
fail to occur for an arbitrary choice of couplings between the fields.
 Thus a generic
theory in $AdS$ would not give a boundary theory which would possess 
a convergent
local OPE. 

In the $\< O_CO_CO_CO_C\>$ 
correlator we found a cancellation among 3 $\phi$-exchange graphs which each
have  a log singularity. The t-channel graviton exchange diagram in this correlator 
is
the same as  the t-channel graviton exchange in $\< O_{\phi} O_CO_{\phi}O_C\>$. 
Suppose
 that this latter graph does contain the cancelling logarithms
discussed above. It is then a simple consequence of \ref{nine} and \ref{ten} that the sum
 of log
 singularities
in the t,s, and u channel  graviton exchange diagrams will also cancel in
  $\< O_CO_CO_CO_C\>$.

Although we  have not evaluated the graviton exchange graphs in Figs. 1 and 2, 
it does appear on physical grounds that they are non-vanishing and have a strong
 singularity $\sim 1/x^4$ for $x \rightarrow 0$, where $x$ is the separation
of any two boundary operators connected to the same internal vertex.
 Part of this physical intuition stems from the fact that the 3-point functions 
$\<O_{C}(x_1) O_{C}(x_2) T_{i j}(x_3)\>$ and 
$\<O_{\phi}(x_1) O_{\phi}(x_2) T_{i j}(x_3)\>$, where
$T_{i j}$ is the stress-energy tensor, 
are different from zero~\cite{tseytlinfirst}, so that we expect
from the leading term of the OPE the singularity $\sim 1/x^{\Delta_1 + \Delta_2 -
\Delta_3}$, where all $\Delta_i =4$. This would 
 imply that the $t$-channel graph in Fig.1 is more singular 
as $x_{13} \rightarrow 0$
 than any of the other graphs, so that the overall sum of all diagrams
contributing to $\< O_{\phi} O_CO_{\phi}O_C\>$ is not expected to vanish.
One can state the same physical expectation in the language of the boundary
 ${\cal N}=4 ~SYM$ theory,
in which $O_\phi = TrF^2$ and $O_C = TrF\tilde F$, and the 2- and 3-point
 functions of these
operators are exactly given by their free-field values due to superconformal
 non-renormalization
theorems. It is easy to calculate the free field OPE's and see that
 $Tr F^2(x) Tr F^2(y)$ and
$TrF\tilde F(x)TrF\tilde F(y)$ contain the
stress tensor with expected $1/(x-y)^4$ singularity. Thus physical considerations
 within the boundary
CFT lead us to expect a non-vanishing t-channel contribution 
to  $\< O_{\phi} O_CO_{\phi}O_C\>$.

It is also easy to understand on physical grounds why the naively expected
 $1/(x_{12})^4$ singularity
of the s-channel graph for $\< O_{\phi} O_CO_{\phi}O_C\>$ is not present.
 First, one can use the
formulae of [5] to show that $\<O_{\phi} O_C O_C \> = 0$ (The AdS integral 
 $\int \frac{d^5 z}{z_0^5} z_0^2 K \partial_{z_{\mu}} K\partial_{z_{\mu}} K $ 
vanishes
even though the action (\ref{one}) contains the vertex $\phi (\partial C)^2$.)
 Second,
one can compute the free field OPE $TrF^2(x) Tr F\tilde F(y)$ and see that
 there is no
$1/(x-y)^4$ singularity (although we expect a weaker singularity from  operators
of dimension greater than 4).

We  comment on the relation between supergravity graphs and OPE's. Consider
a 4-point correlator of chiral primaries, $\< O_1(x_1) O_2(x_2) O_3(x_3) O_4(x_4)\>$.
In the expansion (\ref{doubleope}), let us consider the sum over chiral
primaries  and their conformal descendents. The $SO(6)$ symmetry
 of the $N=4 ~SYM$ theory 
allows only a finite number
of chiral primaries to appear in this expansion. The same symmetry of the
 $AdS_5\times S_5$ supergravity theory
 allows
only a finite number of fields to propagate in the internal
lines of the corresponding $AdS$ graphs. It is thus tempting to seek a relation
between, say,  the s-channel $AdS$ graph whose internal line corresponds to a
 specific primary operator $O(x)$
and the contribution of $O(x)$ and its descendents (i.e. derivatives) in the double
 OPE (23). Consider the limit
$x_{12}$ small, $x_{34}$ small, $x_{13}$ large. The $s$--channel supergravity graph
has two 3-point vertices in the interior of $AdS$. Generically, we expect
 large contributions from two distinct domains of integration in the
 space of $z$ and $w$:
(a) $z$ is near $\vec{x}_1$,$\vec{x}_2$, while $w$ is near
$\vec{x}_3$,$\vec{x}_4$; (b) both $z$ and $w$ are near
$\vec{x}_1$,$\vec{x}_2$ (or both near $\vec{x}_3$, $\vec{x}_4$).

In  region (a) the bulk supergravity propagator goes from near
one pair to near the other pair, so this contribution might correspond to the double
 OPE (\ref{doubleope}). 
A toy example to study this hypothesis 
was presented in~\cite{talk}. The $CFT$ and $AdS$ calculations were 
compared to fourth
order in $\frac{x_{12}}{x_{13}}$ and $\frac{x_{34}}{x_{13}}$, 
and exact agreement was obtained. 
Recently, in \cite{tseytlin} it was argued that a generic $s$--channel
 supergravity
graph exactly matches the corresponding OPE contribution. However the
 argument relied
on an implicit assumption of analyticity (in order to separate terms
 with physical and
shadow singularities) 
which is not satisfied if there are logarithmic singularities.
Thus the identification of $s$-channel graphs and double $OPE$ contributions 
may not
be exact.  For example,  since the 3-point function $\<O_{\phi} O_C O_C
\>$ vanishes,
the double OPE for the correlator $\< O_{\phi} O_C O_{\phi} O_C \> $ would
 also be naively expected
to vanish. However, we showed
explicitly in Section 3
that the corresponding supergravity $s$-channel graph (Fig.1,s) has a
leading singularity
which is logarithmic. It is an important problem for future work to
determine the exact 
circumstances under which logarthmic singularities occur. This will
require detailed input
from the $AdS_5\times S_5$ bulk supergravity theory, since $s$-channel graphs
formed from derivative
and non-derivative $\phi^3$ vertices may have different analyticity
properties.

We finally would like to make some comments on the issues
of duality both on the supergravity and the CFT side. Supergravity graphs are
not expected to be dual, indeed in the $\phi C \phi C$ example we found that
the $s$ and $u$ channels are manifestly different since they exhibit different
singularities. Operator product expansions are instead dual by definition
under the assumption of their convergence. It appears unlikely that
 ${\cal N}=4$, $d=4$ $SU(N)$ SYM in the $N \rightarrow
\infty$, $g_{YM}^2 N \rightarrow \infty$ limit possesses 
a convergent OPE in terms
of only chiral primaries and their descendents,
if one assumes the validity of the AdS/CFT correspondence. Consider again
$\< O_{\phi}(x_1) O_C(x_2) O_{\phi}(x_3) O_C(x_4) \> $.
The only chiral primary that could enter the double OPE (\ref{doubleope}) is
$O_C$, but the coupling is zero since 
$\<O_{\phi} O_C O_C \> = 0$. Hence in this way of doing the OPE we expect a 
zero answer from the chiral sector. However, using the OPE to expand
$O_{\phi}(x_1) O_{\phi}(x_3)$ and $O_C(x_2) O_C(x_4)$, only the stress-energy
tensor $T_{ij}$ can enter as an intermediate chiral operator, and the
coupling is this time non-zero since $\<O_{\phi} O_{\phi} T_{ij}\>$
and $\<O_{C} O_{C} T_{ij}\>$ do not vanish as shown in~\cite{tseytlinfirst}. 
We thus see that
the assumption of a convergent OPE in terms of only chiral operators appears to
lead to a contradiction. It would be interesting to find out the minimum set of
 operators needed in the theory to allow duality of the OPE expansion for chiral field correlators.

\bigskip

\section*{Acknowledgements}
It is a pleasure to thank E. D'Hoker, S. Ferrara, S. Gubser, I. Klebanov, J. Maldacena
and A. Zaffaroni
 for helpful discussions. 
The research of D.Z.F. is supported in part by
NSF Grant No. PHY-97-22072, S.D.M., A.M. and L.R. by D.O.E. cooperative agreement
DE-FC02-94ER40818. L.R. is  supported in part by INFN `Bruno Rossi' Fellowship.

\end{document}